\documentstyle[12pt]{article}
\def\be {\begin{equation}}
\def\ee {\end{equation}}
\def\ba {\begin{eqnarray}}
\def\ea {\end{eqnarray}}

\def\bi {\begin{itemize}}
\def\ei {\end{itemize}}
\begin{document}
\def\bea{\begin{eqnarray}}
\def\eea{\end{eqnarray}}
\title{\bf  {Interacting  Holographic Phantom}}
 \author{M.R. Setare  \footnote{E-mail: rezakord@ipm.ir}
  \\{Department of Science,  Payame Noor University. Bijar. Iran}}
\date{\small{}}

\maketitle
\begin{abstract}
In this paper we consider the holographic model of interacting dark
energy in non-flat universe. With the choice of $c\leq 0.84$, the
interacting holographic dark energy can be described  by a phantom
scalar field. Then we show this phantomic description of the
holographic dark energy with $c\leq 0.84$ and reconstruct the
potential of the phantom scalar field.

 \end{abstract}

\newpage

\section{Introduction}
Nowadays it is strongly believed that the universe is experiencing
an accelerated expansion. Recent observations from type Ia
supernovae \cite{SN} in associated with Large Scale Structure
\cite{LSS} and Cosmic Microwave Background anisotropies \cite{CMB}
have provided main evidence for this cosmic acceleration. In order
to explain why the cosmic acceleration happens, many theories have
been proposed. Although theories of trying to modify Einstein
equations constitute a big part of these attempts, the mainstream
explanation for this problem, however, is known as theories of dark
energy. It is the most accepted idea that a mysterious dominant
component, dark energy, with negative pressure, leads to this cosmic
acceleration, though its nature and cosmological origin still remain
enigmatic at present.

The combined analysis of cosmological observations suggests that the
universe consists of about $70\%$ dark energy, $30\%$ dust matter
(cold dark matter plus baryons), and negligible radiation. Although
the nature and origin of dark energy are unknown, we still can
propose some candidates to describe it. The most obvious theoretical
candidate of dark energy is the cosmological constant $\lambda$ (or
vacuum energy) \cite{Einstein:1917,cc} which has the equation of
state $w=-1$. However, as is well known, there are two difficulties
arise from the cosmological constant scenario, namely the two famous
cosmological constant problems --- the ``fine-tuning'' problem and
the ``cosmic coincidence'' problem \cite{coincidence}.

 An alternative proposal for dark energy is the
dynamical dark energy scenario. The cosmological constant puzzles
may be better interpreted by assuming that the vacuum energy is
canceled to exactly zero by some unknown mechanism and introducing a
dark energy component with a dynamically variable equation of state.
The dynamical dark energy proposal is often realized by some scalar
field mechanism which suggests that the energy form with negative
pressure is provided by a scalar field evolving down a proper
potential. So far, a large class of scalar-field dark energy models
have been studied, including quintessence \cite{quintessence},
K-essence \cite{kessence}, tachyon \cite{tachyon}, phantom
\cite{phantom}, ghost condensate \cite{ghost1,ghost2} and quintom
\cite{quintom}, and so forth. But we should note that the mainstream
viewpoint regards the scalar field dark energy models as an
effective description of an underlying theory of dark energy. In
addition, other proposals on dark energy include interacting dark
energy models \cite{intde}, braneworld models \cite{brane}, and
Chaplygin gas models \cite{cg}, etc.. One should realize,
nevertheless, that almost these models are settled at the
phenomenological level, lacking theoretical root.

It is generally believed by theorists that we can not entirely
understand the nature of dark energy before a complete theory of
quantum gravity is established \cite{Witten:2000zk}. However,
although we are lacking a quantum gravity theory today, we still can
make some attempts to probe the nature of dark energy according to
some principles of quantum gravity. The holographic dark energy
model is just an appropriate example, which is constructed in the
light of the holographic principle of quantum gravity theory. That
is to say, the holographic dark energy model possesses some
significant features of an underlying theory of dark energy.

The distinctive feature of the cosmological constant or vacuum
energy is that its equation of state is always exactly equal to
$-1$. However, when considering the requirement of the holographic
principle originating from the quantum gravity speculation, the
vacuum energy will acquire dynamically property. As we speculate,
the dark energy problem may be in essence a problem belongs to
quantum gravity \cite{Witten:2000zk}. In the classical gravity
theory, one can always introduce a cosmological constant to make the
dark energy density be an arbitrary value. However, a complete
theory of quantum gravity should be capable of making the properties
of dark energy, such as the energy density and the equation of
state, be determined definitely and uniquely. Currently, an
interesting attempt for probing the nature of dark energy within the
framework of quantum gravity is the so-called ``holographic dark
energy'' proposal
\cite{Cohen:1998zx,Horava:2000tb,Hsu:2004ri,Li:2004rb}. It is well
known that the holographic principle is an important result of the
recent researches for exploring the quantum gravity (or string
theory) \cite{holoprin}. This principle is enlightened by
investigations of the quantum property of black holes. Roughly
speaking, in a quantum gravity system, the conventional local
quantum field theory will break down. The reason is rather simple:
For a quantum gravity system, the conventional local quantum field
theory contains too many degrees of freedom, and such many degrees
of freedom will lead to the formation of black hole so as to break
the effectiveness of the quantum field theory.

For an effective field theory in a box of size $L$, with UV cut-off
$\Lambda$ the entropy $S$ scales extensively, $S\sim L^3\Lambda^3$.
However, the peculiar thermodynamics of black hole \cite{bh} has led
Bekenstein to postulate that the maximum entropy in a box of volume
$L^3$ behaves nonextensively, growing only as the area of the box,
i.e. there is a so-called Bekenstein entropy bound, $S\leq
S_{BH}\equiv\pi M_P^2L^2$. This nonextensive scaling suggests that
quantum field theory breaks down in large volume. To reconcile this
breakdown with the success of local quantum field theory in
describing observed particle phenomenology, Cohen et al.
\cite{Cohen:1998zx} proposed a more restrictive bound -- the energy
bound. They pointed out that in quantum field theory a short
distance (UV) cut-off is related to a long distance (IR) cut-off due
to the limit set by forming a black hole. In other words, if the
quantum zero-point energy density $\rho_{\Lambda}$ is relevant to a
UV cut-off $\Lambda$, the total energy of the whole system with size
$L$ should not exceed the mass of a black hole of the same size,
thus we have $L^3\rho_{\Lambda}\leq LM_P^2$. This means that the
maximum entropy is in order of $S_{BH}^{3/4}$. When we take the
whole universe into account, the vacuum energy related to this
holographic principle \cite{holoprin} is viewed as dark energy,
usually dubbed holographic dark energy. The largest IR cut-off $L$
is chosen by saturating the inequality so that we get the
holographic dark energy density
\begin{equation}
\rho_{\Lambda}=3c^2M_P^2L^{-2}~,\label{de}
\end{equation} where $c$ is a numerical constant, and $M_P\equiv 1/\sqrt{8\pi
G}$ is the reduced Planck mass. Many authors have devoted to
developed the idea of the holographic dark energy. It has been
demonstrated that it seems most likely that the IR cutoff is
relevant to the future event horizon
\begin{equation}
R_{\rm h}(a)=a\int_t^\infty{dt'\over a(t')}=a\int_a^\infty{da'\over
Ha'^2}~.\label{eh}
\end{equation} Such a holographic dark energy looks reasonable, since
it may provide simultaneously natural solutions to both dark energy
problems as demonstrated in Ref.\cite{Li:2004rb}. The holographic
dark energy model has been tested and constrained by various
astronomical observations \cite{obs1,obs2,obs3}. Furthermore, the
holographic dark energy model has been extended to include the
spatial curvature contribution, i.e. the holographic dark energy
model in non-flat space \cite{nonflat}. We focus in this paper on
the holographic dark energy in a non-flat universe. For other
extensive studies, see e.g. \cite{holoext}.\\
It is known that the coincidence or "why now" problem is easily
solved in some models of HDE based on this fundamental assumption
that matter and holographic dark energy do not conserve separately,
but the matter energy density decays into the holographic energy
density \cite{interac}. In fact a suitable evolution of the Universe
is obtained when, in addition to the holographic dark energy, an
interaction (decay of
dark energy to matter) is assumed.\\
In the present paper, we suggest  a correspondence between the
holographic dark energy scenario and the phantom dark energy model.
The current available observational data imply that the holographic
vacuum energy behaves as phantom-type dark energy, i.e. the
equation-of-state of dark energy crosses the cosmological-constant
boundary $w=-1$ during the evolution history. We show this phantomic
description of the interacting holographic dark energy in non-flat
universe with $c\leq 0.84$, and reconstruct the potential of the
phantom scalar field.

\section{ Interacting holographic phantom in non-flat universe }
In this section we obtain the equation of state for the holographic
energy density when there is an interaction between holographic
energy density $\rho_{\Lambda}$ and a Cold Dark Matter(CDM) with
$w_{m}=0$. The continuity equations for dark energy and CDM are
\begin{eqnarray}
\label{2eq1}&& \dot{\rho}_{\rm \Lambda}+3H(1+w_{\rm \Lambda})\rho_{\rm \Lambda} =-Q, \\
\label{2eq2}&& \dot{\rho}_{\rm m}+3H\rho_{\rm m}=Q.
\end{eqnarray}
The interaction is given by the quantity $Q=\Gamma \rho_{\Lambda}$.
This is a decaying of the holographic energy component into CDM with
the decay rate $\Gamma$. Taking a ratio of two energy densities as
$u=\rho_{\rm m}/\rho_{\rm \Lambda}$, the above equations lead to
\begin{equation}
\label{2eq3} \dot{u}=3Hu\Big[w_{\rm \Lambda}+
\frac{1+u}{u}\frac{\Gamma}{3H}\Big]
\end{equation}
 Following Ref.\cite{Kim:2005at},
if we define
\begin{eqnarray}\label{eff}
w_\Lambda ^{\rm eff}=w_\Lambda+{{\Gamma}\over {3H}}\;, \qquad w_m
^{\rm eff}=-{1\over u}{{\Gamma}\over {3H}}\;.
\end{eqnarray}
Then, the continuity equations can be written in their standard
form
\begin{equation}
\dot{\rho}_\Lambda + 3H(1+w_\Lambda^{\rm eff})\rho_\Lambda =
0\;,\label{definew1}
\end{equation}
\begin{equation}
\dot{\rho}_m + 3H(1+w_m^{\rm eff})\rho_m = 0\; \label{definew2}
\end{equation}
We consider the non-flat Friedmann-Robertson-Walker universe with
line element
 \be\label{metr}
ds^{2}=-dt^{2}+a^{2}(t)(\frac{dr^2}{1-kr^2}+r^2d\Omega^{2}).
 \ee
where $k$ denotes the curvature of space k=0,1,-1 for flat, closed
and open universe respectively. A closed universe with a small
positive curvature ($\Omega_k\sim 0.01$) is compatible with
observations \cite{ {wmap}, {ws}}. We use the Friedmann equation to
relate the curvature of the universe to the energy density. The
first Friedmann equation is given by
\begin{equation}
\label{2eq7} H^2+\frac{k}{a^2}=\frac{1}{3M^2_p}\Big[
 \rho_{\rm \Lambda}+\rho_{\rm m}\Big].
\end{equation}
In non-flat universe, our choice for holographic dark energy density
is
 \be \label{holoda}
  \rho_\Lambda=3c^2M_{p}^{2}L^{-2}.
 \ee
$L$ is defined as the following form\cite{nonflat}:
\begin{equation}\label{leq}
 L=ar(t),
\end{equation}
here, $a$, is scale factor and $r(t)$ is relevant to the future
event horizon of the universe. Given the fact that
\begin{eqnarray}
\int_0^{r_1}{dr\over \sqrt{1-kr^2}}&=&\frac{1}{\sqrt{|k|}}{\rm
sinn}^{-1}(\sqrt{|k|}\,r_1)\nonumber\\
&=&\left\{\begin{array}{ll}
\sin^{-1}(\sqrt{|k|}\,r_1)/\sqrt{|k|},\ \ \ \ \ \ &k=1,\\
r_1,&k=0,\\
\sinh^{-1}(\sqrt{|k|}\,r_1)/\sqrt{|k|},&k=-1,
\end{array}\right.
\end{eqnarray}
one can easily derive \be \label{leh} L=\frac{a(t) {\rm
sinn}[\sqrt{|k|}\,R_{h}(t)/a(t)]}{\sqrt{|k|}},\ee where $R_h$ is the
future event horizon given by (\ref{eh}). By considering  the
definition of holographic energy density $\rho_{\rm \Lambda}$, one
can find \cite{{set1},{set2}}:
\begin{equation}\label{stateq}
w_{\rm \Lambda}=-[\frac{1}{3}+\frac{2\sqrt{\Omega_{\rm
\Lambda}}}{3c}\frac{1}{\sqrt{|k|}}\rm
cosn(\sqrt{|k|}\,R_{h}/a)+\frac{\Gamma}{3H}].
\end{equation}
 where
\begin{equation}
\frac{1}{\sqrt{|k|}}{\rm cosn}(\sqrt{|k|}x)
=\left\{\begin{array}{ll}
\cos(x),\ \ \ \ \ \ &k=1,\\
1,&k=0,\\
\cosh(x),&k=-1.
\end{array}\right.
\end{equation}
Here as in Ref.\cite{WGA}, we choose the following relation for
decay rate
\begin{equation}\label{decayeq}
\Gamma=3b^2(1+u)H
\end{equation}
with  the coupling constant $b^2$. Substitute this relation into
Eq.(\ref{stateq}), one finds the holographic energy equation of
state \cite{set1}
\begin{equation} \label{3eq4}
w_{\rm \Lambda}=-\frac{1}{3}-\frac{2\sqrt{\Omega_{\rm
\Lambda}-c^2\Omega_{k}}}{3c}-\frac{b^2(1+\Omega_{k})}{\Omega_{\rm
\Lambda}}.
\end{equation}
From Eqs.(\ref{eff}, \ref{3eq4}), we have the effective equation of
state as \footnote{It seems that $w_{\rm \Lambda}^{eff}$ is
independent of the coupling constant $b$, however, a solution
$\Omega_{\Lambda}$ to the following evolution equation which
includes the the $b^2-$terms determines how the effective equation
of state $w_{\rm \Lambda}^{eff}$ is changing under the evolution of
the universe. The differential equation for $\Omega_{\Lambda}$ is
\be \label{evol}
\frac{d\Omega_{\Lambda}}{dx}=\frac{\dot{\Omega_{\Lambda}}}{H}=3\Omega_{\Lambda}(1+\Omega_{k}-\Omega_{\Lambda})
[-\frac{1}{3}-\frac{2\sqrt{\Omega_{\rm
\Lambda}}}{3c}\frac{1}{\sqrt{|k|}}\rm
cosn(\sqrt{|k|}\,R_{h}/a)-b^2\frac{(1+\Omega_{k})}{\Omega_{\Lambda}}+\frac{b^2(1+\Omega_{k})^{2}}
{(1+\Omega_{k}-\Omega_{\Lambda})\Omega_{\Lambda}}] \ee}
\begin{equation} \label{3eq401}
w_{\rm \Lambda}^{eff}=-\frac{1}{3}-\frac{2\sqrt{\Omega_{\rm
\Lambda}-c^2\Omega_{k}}}{3c}.
\end{equation}
 For the non-flat universe, the authors of \cite{obsnonflat} used the data
coming from the SN and CMB to constrain the holographic dark energy
model, and got the 1 $\sigma$ fit results: $c=0.84^{+0.16}_{-0.03}$.
If we take $c=0.84$, and taking $\Omega_{\Lambda}=0.73$,
$\Omega_{k}=0.01$ for the present time, using Eq.(\ref{3eq401}) we
obtain $w_{\rm \Lambda}^{eff}=-1.007$. Also for the flat case, the
X-ray gas mass fraction of rich clusters, as a function of redshift,
has also been used to constrain the holographic dark energy model
\cite{obs2}. The main results, i.e. the 1 $\sigma$ fit values for
$c$ is: $c=0.61^{+0.45}_{-0.21}$, in this case also we obtain
$w_{\rm \Lambda}^{eff}<-1$. This implies that one can generate
phantom-like equation of state from an interacting holographic dark
energy model in flat and
non-flat universe only if $c\leq 0.84$. This upper limit on $c$ depends on the exact value of
$\Omega_{\Lambda}$ and therefore depend on the coupling constant $b$ (see Eq.(\ref{evol}). Also
it depend on the value of $\Omega_{k}$ and cutoff $L$.  \\
Now we assume that the origin of the dark energy is a phantom scalar
field $\phi$, so \be \label{roph1}
\rho_{\Lambda}=-\frac{1}{2}\dot{\phi}^{2}+V(\phi) \ee \be
\label{roph2} P_{\Lambda}=-\frac{1}{2}\dot{\phi}^{2}-V(\phi) \ee In
this case $w_{\rm \Lambda}$ is given by \be \label{w}w_{\rm
\Lambda}=\frac{-\frac{1}{2}\dot{\phi}^{2}-V(\phi)}{-\frac{1}{2}\dot{\phi}^{2}+V(\phi)}
\ee One can see that in this case $w_{\rm \Lambda}<-1$
\footnote{Differenating Eq.(\ref{2eq7}) with respect to the cosmic
time $t$, one find \be \label{hdot}\dot{H}=\frac{\dot{\rho}}{6H
M_{p}^{2}}+\frac{k}{a^2} \ee where $\rho=\rho_{m}+\rho_{\Lambda}$ is
the total energy density, now using Eqs.(\ref{2eq1}, \ref{2eq2}) \be
\label{doro} \dot{\rho}=-3H(1+w)\rho \ee  where \be
\label{weq}w=\frac{w_{\Lambda}\rho_{\Lambda}}{\rho}=\frac{\Omega_{\Lambda}w_{\Lambda}}{1+\frac{k}{a^2H^2}}
\ee Substitute $\dot{\rho}$ into Eq.(\ref{hdot}), we obtain \be
\label{weq2}
w=\frac{2/3(\frac{k}{a^2}-\dot{H})}{H^2+\frac{k}{a^2}}-1 \ee In a
phantom dominated universe $\dot{H}>0$, from Eq.(\ref{weq2}) one can
see easily that in the $k=0, k=-1$ cases $w<-1$, therefore in this
cases $w_{\Lambda}<-1$ also. For $k=1$, the necessary condition to
obtain $w_{\Lambda}<-1$ is this: $\dot{H}>\frac{1}{a^2}$. However as
we have shown above, with the choice of $c\leq 0.84$, the
holographic dark energy model can predict a phantom era,  even for
$k=1$ case.}, therefore, only
 holographic dark energy in cases $c\leq 0.84$ can be described by the
phantom. It must be pointed out that the choice of $c\leq 0.84$, on
theoretical level, will bring some troubles. The Gibbons-Hawking
entropy will thus decrease since the event horizon shrinks, which
violates the second law of thermodynamics as well. However, the
current observational data indicate that the parameter $c$ in the
holographic model seems smaller than 1. Now we reconstruct the
phantom potential and the dynamics of the scalar field in light of
the holographic dark energy with $c\leq 0.84$. According to the
forms of phantom energy density and pressure eqs.(\ref{roph1},
\ref{roph2}), one can easily derive the scalar potential and kinetic
energy term as \be \label{v} V(\phi)=\frac{1}{2}(1-w_{\rm
\Lambda})\rho_{\Lambda} \ee \be \label{phi}\dot{\phi}^{2}
=-(1+w_{\rm \Lambda})\rho_{\Lambda} \ee Using Eqs.(\ref{weq},
\ref{weq2}), one can rewrite the holographic energy equation of
state as \be \label{eqes1}
w_{\Lambda}=\frac{-1}{3\Omega_{\Lambda}H^{2}}(2\dot{H}+3H^2+\frac{k}{a^2})
\ee Substitute the above $w_{\Lambda}$ into Eqs.(\ref{v},
\ref{phi}), we obtain \be \label{v1} V(\phi)=\frac{M_{p}^{2}}{2}
[2\dot{H}+3H^2(1+\Omega_{\Lambda})+\frac{k}{a^2}]\ee  \be
\label{phi2}\dot{\phi}^{2}
=M_{p}^{2}[2\dot{H}+3H^2(1-\Omega_{\Lambda})+\frac{k}{a^2}]\ee In
the spatially flat case, $k=0$,and  $\Omega_{\Lambda}=1$, in this
case the Eqs.(\ref{v1}, \ref{phi2}) are exactly Eq.(6) in \cite
{odi1} if we consider $\omega(\phi)=-1$. In similar to the \cite
{{odi1}, {odi2}}, we can define $\dot{\phi}^{2}$ and $V(\phi)$ in
terms of single function $f(\phi)$ as \be \label{v2}
V(\phi)=\frac{M_{p}^{2}}{2}
[2f'(\phi)+3f^{2}(\phi)(1+\Omega_{\Lambda})+\frac{k}{a^2}]\ee \be
\label{phi3}1=M_{p}^{2}
[2f'(\phi)+3f^{2}(\phi)(1-\Omega_{\Lambda})+\frac{k}{a^2}]\ee In the
spatially flat case the Eqs.(\ref{v2}, \ref{phi3}) solved only in
case of presence of two scalar potentials  $V(\phi)$, and
$\omega(\phi)$. Here we have claimed that in the presence of
curvature term $\frac{k}{a^2}$, Eqs.(\ref{v2}, \ref{phi3}) may be
solved with  potential  $V(\phi)$ ( To see the general procedure for
such type calculations refer to \cite{{odi1},{odi2}}, see also the
appendix of the present paper). Hence, the following solution are
obtained \be \label{sol} \phi=t, \hspace{1cm} H=f(t) \ee One can
check that the solution (\ref{sol}) satisfies the following scalar
field equation \be \label{phieq}-\ddot{\phi}-3H\dot{\phi}+V'(\phi)=0
\ee Therefore by the above condition, $f(\phi)$ in our model must
satisfy following relation \be \label{coneq} 3f(\phi)=V'(\phi)\ee
 In the other hand, using
Eqs.(\ref{holoda}, \ref{3eq4}) we have \be
\label{v4}V(\phi)=\frac{3H^2 \Omega_{\rm \Lambda}}{16\pi
G}(\frac{4}{3}+\frac{2\sqrt{\Omega_{\rm
\Lambda}-c^2\Omega_{k}}}{3c}+\frac{b^2(1+\Omega_k)}{\Omega_{\rm
\Lambda}}) \ee \be \label{phi1} \dot{\phi}=\frac{H\sqrt{\Omega_{\rm
\Lambda}}}{2\sqrt{\pi G}}[-1+\frac{\sqrt{\Omega_{\rm
\Lambda}-c^2\Omega_{k}}}{c}+\frac{3b^2(1+\Omega_k)}{2\Omega_{\rm
\Lambda}}]^{1/2} \ee Using Eq.(\ref{phi1}), we can rewrite
Eq.(\ref{v4}) as \be \label{v5}V(\phi)=3 M_{p}^{2}
H^2(1+\frac{\dot{\phi}^{2}}{6M_{p}^{2} H^2 \Omega_{\rm
\Lambda}}),\ee or in another form as following \be
\label{v5}V(\phi)=3 M_{p}^{2} [f^{2}(\phi)+\frac{1}{6M_{p}^{2}
 \Omega_{\rm \Lambda}}]\ee Then, from Eqs.(\ref{v2},
\ref{v5}), we get \be
\label{keq}\frac{k}{a^{2}}=3f^{2}(\phi)(1-\Omega_{\rm
\Lambda})-2f'(\phi)+\frac{1}{\Omega_{\rm \Lambda}M_{p}^{2}} \ee Now,
using Eqs.(\ref{phi3}, \ref{keq}) we obtain following second order
equation for $\Omega_{\rm \Lambda}$ \be \label{omegaeq}
6M_{p}^{2}f^2 \Omega_{\rm \Lambda}^{2}+(1-6M_{p}^{2}f^2) \Omega_{\rm
\Lambda}-1=0\ee The solutions of this equation are as \be
\label{sol1} \Omega_{\rm
\Lambda}=\frac{(6M_{p}^{2}f^2-1)\pm\sqrt{(1-6M_{p}^{2}f^2)^{2}+24M_{p}^{2}f^2
}}{12M_{p}^{2}f^2}\ee Substitute the above $\Omega_{\rm \Lambda}$
into Eq.(\ref{v5}), we obtain the scalar potential as following\be
\label{v6}V(\phi)=3 M_{p}^{2}f^{2}(\phi)
[1+\frac{2}{(6M_{p}^{2}f^2(\phi)-1)\pm\sqrt{(1-6M_{p}^{2}f^2(\phi))^{2}+24M_{p}^{2}f^2(\phi)
}}] \ee
 \section{Conclusions}
Based on cosmological state of holographic principle, proposed by
Fischler and Susskind \cite{fischler}, the Holographic model of
Dark Energy (HDE) has been proposed and studied widely in the
 literature \cite{Li:2004rb,HDE}. In \cite{HG} using the type Ia
 supernova data, the model of HDE is constrained once
 when c is unity and another time when c is taken as free
 parameter. It is concluded that the HDE is consistent with recent observations, but future observations are needed to
 constrain this model more precisely. In another paper \cite{HL},
 the anthropic principle for HDE is discussed. It is found that,
 provided that the amplitude of fluctuation are variable the
 anthropic consideration favors the HDE over the cosmological
 constant. For flat universe the convenient horizon looks to be $R_h$ while in
non-flat universe we define $L$ because of the problems that arise
if we consider $R_h$ or $R_p$ (particle horizon) \cite{easther}. As
it was mentioned in introduction, $c$ is a positive constant in
holographic model of dark energy, and($c\geq1$). However, if $c<1$,
the holographic dark energy will behave like a phantom model of DE,
the amazing feature of which is that the equation of state of dark
energy component $w_{\rm \Lambda}$ crosses $-1$. Hence, we see, the
determining of the value of $c$ is a key point to the feature of the
holographic dark energy and the ultimate fate of the universe as
well. However, in the recent fit studies, different groups gave
different values to $c$. A direct fit of the present available SNe
Ia data with this holographic model indicates that the best fit
result is $c=0.21$ \cite{HG}. Recently, by calculating the average
equation of state of the dark energy and the angular scale of the
acoustic oscillation from the BOOMERANG and WMAP data on the CMB to
constrain the holographic dark energy model, the authors show that
the reasonable result is $c\sim 0.7$ \cite{cmb1}. In the other hand,
in the study of the constraints on the dark energy from the
holographic connection to the small $l$ CMB suppression, an opposite
result is derived, i.e. it implies the best fit result is $c=2.1$
\cite{cmb3}. Also, the authors of \cite{obsnonflat} used the data
coming from the SN and CMB to constrain the holographic dark energy
model, and got the 1 $\sigma$ fit results: $c=0.84^{+0.16}_{-0.03}$.\\

In this paper we have associated the interacting holographic dark
energy in non-flat universe with a phantom scalar field. We have
shown that the holographic dark energy with $c\leq 0.84$ (here we
have assumed $c$ is a constant, to see similar calculation where $c
$ is variable with time refere to \cite {zim}) can be described
 by the phantom in a certain way. In the another term our calculation show, taking $\Omega_{\Lambda}=0.73$ for
the present time, the upper bound of $c$ is $0.84$. Then a
correspondence between the holographic dark energy and phantom has
been established, and the potential of the holographic phantom has
been reconstructed. This results is in contrast with some other
variants of holographic dark energy discussed in the literature, for
example: The authors of \cite{Kim:2005at} have shown that an
interacting holographic dark energy model cannot accommodate a
transition from the dark energy with $w_{\Lambda}^{eff}\geq -1$ to
the phantom regim with $w_{\Lambda}^{eff}< -1$, also in \cite{setbd}
I have shown that one can not generate phantom-like equation of
state from an interacting holographic dark energy model in non-flat
universe in the Brans-Dicke cosmology framework. In the other hand,
in \cite{setbd} I have suggested a correspondence between the
holographic dark energy scenario in flat universe and the phantom
dark energy model in framework of Brans-Dicke theory with potential.
\section{Acknowledgment}
The author would like to thank the referee because of his useful
comments, which assisted to prepare better frame for this study.
\section{Appendix}
It has been shown in \cite{{odi1},{odi2}} that ark energy dynamics
of the universe can be achieved by equivalent mathematical
descriptions taking into account generalized fluid equations of
state in General Relativity, scalar-tensor theories or modified $F(R)$ gravity in Einstein or Jordan frames.\\
We consider following action \be \label{a1}S=\int\sqrt{-g} d^{4}x
[\frac{M_{p}^{2}}{2}R-\frac{1}{2}\omega(\varphi)\partial_{\mu}
\varphi \partial^{\mu}\varphi-\tilde{V}(\varphi)]\ee The energy
density and pressure are given by
 \be \label{a2}
\rho_{\Lambda}=\frac{1}{2}\omega(\varphi)\dot{\varphi}^{2}+\tilde{V}(\varphi)
\ee \be \label{a3}
P_{\Lambda}=\frac{1}{2}\omega(\varphi)\dot{\varphi}^{2}-\tilde{V}(\varphi)
\ee According to the forms of energy density and pressure
eqs.(\ref{a2}, \ref{a3}), one can easily derive the scalar potential
and kinetic energy term as \be \label{a4}
\tilde{V}(\varphi)=\frac{1}{2}(1-w_{\rm \Lambda})\rho_{\Lambda} \ee
\be \label{a5}\omega(\varphi)\dot{\varphi}^{2} =(1+w_{\rm
\Lambda})\rho_{\Lambda} \ee Using Eq.(\ref{eqes1}), one can obtain
\be \label{a6}\tilde{V}(\varphi)=\frac{M_{p}^{2}}{2}
[2\dot{H}+3H^2(1+\Omega_{\Lambda})+\frac{k}{a^2}]\ee  \be
\label{a7}\omega(\varphi)\dot{\varphi}^{2}
=-M_{p}^{2}[2\dot{H}+3H^2(1-\Omega_{\Lambda})+\frac{k}{a^2}]\ee The
interesting case is that $\omega(\varphi)$ and $\tilde{V}(\varphi)$
are defined in terms of a single function $f(\varphi)$ as \be
\label{a8} \tilde{V}(\varphi)=\frac{M_{p}^{2}}{2}
[2f'(\varphi)+3f^{2}(\varphi)(1+\Omega_{\Lambda})+\frac{k}{a^2}]\ee
\be \label{a9}\omega(\varphi)=-M_{p}^{2}
[2f'(\varphi)+3f^{2}(\varphi)(1-\Omega_{\Lambda})+\frac{k}{a^2}]\ee
Hence, the following solution are obtained \be \label{a10}
\varphi=t, \hspace{1cm} H=f(t) \ee
 In the other hand, using
Eqs.(\ref{holoda}, \ref{3eq4}) we have \be
\label{a11}\tilde{V}(\varphi)=\frac{3H^2 \Omega_{\rm \Lambda}}{16\pi
G}(\frac{4}{3}+\frac{2\sqrt{\Omega_{\rm
\Lambda}-c^2\Omega_{k}}}{3c}+\frac{b^2(1+\Omega_k)}{\Omega_{\rm
\Lambda}}) \ee \be \label{a12} \sqrt{|\omega(\varphi) |}
\dot{\varphi}=\frac{H\sqrt{\Omega_{\rm \Lambda}}}{2\sqrt{\pi
G}}[-1+\frac{\sqrt{\Omega_{\rm
\Lambda}-c^2\Omega_{k}}}{c}+\frac{3b^2(1+\Omega_k)}{2\Omega_{\rm
\Lambda}}]^{1/2} \ee Using Eq.(\ref{a12}), we can rewrite
Eq.(\ref{a11}) as \be \label{a13}\tilde{V}(\varphi)= 3 M_{p}^{2}
[f^{2}(\varphi)-\frac{\omega(\varphi)}{6M_{p}^{2}
 \Omega_{\rm \Lambda}}]\ee Then, from Eqs.(\ref{a8},
\ref{a13}), we get \be
\label{a14}\frac{k}{a^{2}}=3f^{2}(\varphi)(1-\Omega_{\rm
\Lambda})-2f'(\varphi)-\frac{\omega(\varphi)}{\Omega_{\rm
\Lambda}M_{p}^{2}} \ee Now, using Eqs.(\ref{a9}, \ref{a14}) we
obtain following second order equation for $\Omega_{\rm \Lambda}$
\be \label{a15} -6M_{p}^{2}f^2 \Omega_{\rm
\Lambda}^{2}+(\omega(\varphi)+6M_{p}^{2}f^2) \Omega_{\rm
\Lambda}-\omega(\varphi)=0\ee The solutions of this equation are as
\be \label{a16} \Omega_{\rm
\Lambda}=\frac{-(6M_{p}^{2}f^2+\omega(\varphi))\pm\sqrt{(\omega(\varphi)+6M_{p}^{2}f^2)^{2}-24M_{p}^{2}f^2\omega(\varphi)
}}{-12M_{p}^{2}f^2}\ee Substitute the above $\Omega_{\rm \Lambda}$
into Eq.(\ref{a13}), we obtain the scalar potential as following\be
\label{a17}\tilde{V}(\varphi)=3 M_{p}^{2}f^{2}(\varphi)
[1+\frac{2\omega(\varphi)}{-(6M_{p}^{2}f^2+\omega(\varphi))\pm\sqrt{(\omega(\varphi)
+6M_{p}^{2}f^2)^{2}-24M_{p}^{2}f^2\omega(\varphi)}}] \ee If we
define a new field $\phi$ as \be \label{a18} \phi= \int d\varphi
\sqrt{|\omega(\varphi)|}\ee the action (\ref{a1}) can be rewritten
as \be \label{a19}S=\int \sqrt{-g} d^{4}x
[\frac{M_{p}^{2}}{2}R\mp\frac{1}{2}\partial_{\mu} \phi
\partial^{\mu}\phi-{V}(\phi)]\ee
The sign in front of the kinetic term depends on the sign of
$\omega(\varphi)$. If the sign of $\omega(\varphi)$  is positive,
the sign of the kinetic term is negative. Therefore, in the phantom
phase, the sign is always positive, this is what we have considered
in this paper. One can assumes $\varphi$ can be solved with respect
to $\phi$: $\varphi=\varphi(\phi)$. Then the potential $V(\phi)$ is
given by $V(\phi)=\tilde{V}(\varphi(\phi))$. We have shown that with
the choice of $c\leq 0.84$ the interacting holographic dark energy
has a phantom-like behaviour. Therefore, the equivalent
scalar-tensor theory for interacting holographic fluid is given by
Eq.(\ref{a19}), where the sign of the kinetic term is negative. In
this paper we did not consider the potential $\omega(\varphi)$, we
have claimed that in the presence of curvature term $\frac{k}{a^2}$,
Eqs.(\ref{v2}, \ref{phi3}) may be solved with potential  $V(\phi)$
only, but with condition $(\ref{coneq})$.

\end{document}